\documentclass[lettersize,journal]{IEEEtran}
\usepackage{amsmath,amsfonts}
\usepackage{algorithmic}
\usepackage{algorithm}
\usepackage{array}
\usepackage[caption=false,font=normalsize,labelfont=sf,textfont=sf]{subfig}
\usepackage{textcomp}
\usepackage{stfloats}
\usepackage{url}
\usepackage{verbatim}
\usepackage{graphicx}
\usepackage{booktabs}
\usepackage{cite}
\usepackage{multirow}
\usepackage{diagbox}
\usepackage[table]{xcolor} 
\usepackage{xcolor}
\usepackage[colorlinks,
            linkcolor=blue,       
            anchorcolor=blue,  
            citecolor=blue,        
            ]{hyperref}
\usepackage{tipa}

\begin{document}

\title{PASE: Phoneme-Aware Speech Encoder to Improve Lip Sync Accuracy for Talking Head Synthesis}

\author{Yihuan Huang, Jiajun Liu, Yanzhen Ren,~\IEEEmembership{Member,~IEEE,}, Jun Xue and Wuyang Liu,~\IEEEmembership{Graduate Student Member,~IEEE,}, Zongkun Sun

\thanks{
--------------------------------------------------------------------------------------------
$^{\ast}$ Yanzhen Ren is the corresponding author.}

\thanks{Yanzhen Ren is with the Key Laboratory of Aerospace Information Security and Trusted Computing, Ministry of Education, School of Cyber Science and Engineering, Wuhan University, Wuhan 430072, China (e-mail: renyz@whu.edu.cn).}
\thanks{Yihuan Huang, Jiajun Liu, Jun Xue, Wuyang Liu are with the School of Cyber Science and Engineering, Wuhan University, Wuhan 430072, China (e-mail:yihuanhuang@whu.edu.cn, jiajunliu@whu.edu.cn, junxue@whu.edu.cn, liuwuyang@whu.edu.cn).}
\thanks{Zongkun Sun is with the School of Police Information, Shandong Police College, Jinan 250200, China. (e-mail: zongksun@sdpc.edu.cn)}
\thanks{
This work is supported by the Natural Science Foundation of China (NSFC) under the grants No. 62572358, 62172306, and 62372334.
}
}



\maketitle

\begin{abstract}
Recent talking head synthesis works typically adopt speech features extracted from large-scale pre-trained acoustic models. However, the intrinsic many-to-many relationship between speech and lip motion causes phoneme–viseme alignment ambiguity, leading to inaccurate and unstable lips. To further improve lip sync accuracy, we propose PASE (Phoneme-Aware Speech Encoder), a novel speech representation model that bridges the gap between phonemes and visemes. PASE explicitly introduces phoneme embeddings as alignment anchors and employs a contrastive alignment module to enhance the discriminability between corresponding audio–visual pairs. In addition, a prediction and reconstruction task is designed to improve robustness under noise and partial modality absence. Experimental results show PASE significantly improves lip sync accuracy and achieves state-of-the-art performance across both NeRF- and 3DGS-based rendering frameworks, outperforming conventional methods based on acoustic features by 13.7\% and 14.2\%, respectively. Importantly, PASE can be seamlessly integrated into diverse talking head pipelines to improve the lip sync accuracy without architectural modifications.
\end{abstract}


\section{Introduction}
\begin{figure*}[t]
  \centering
  \includegraphics[width=\linewidth]{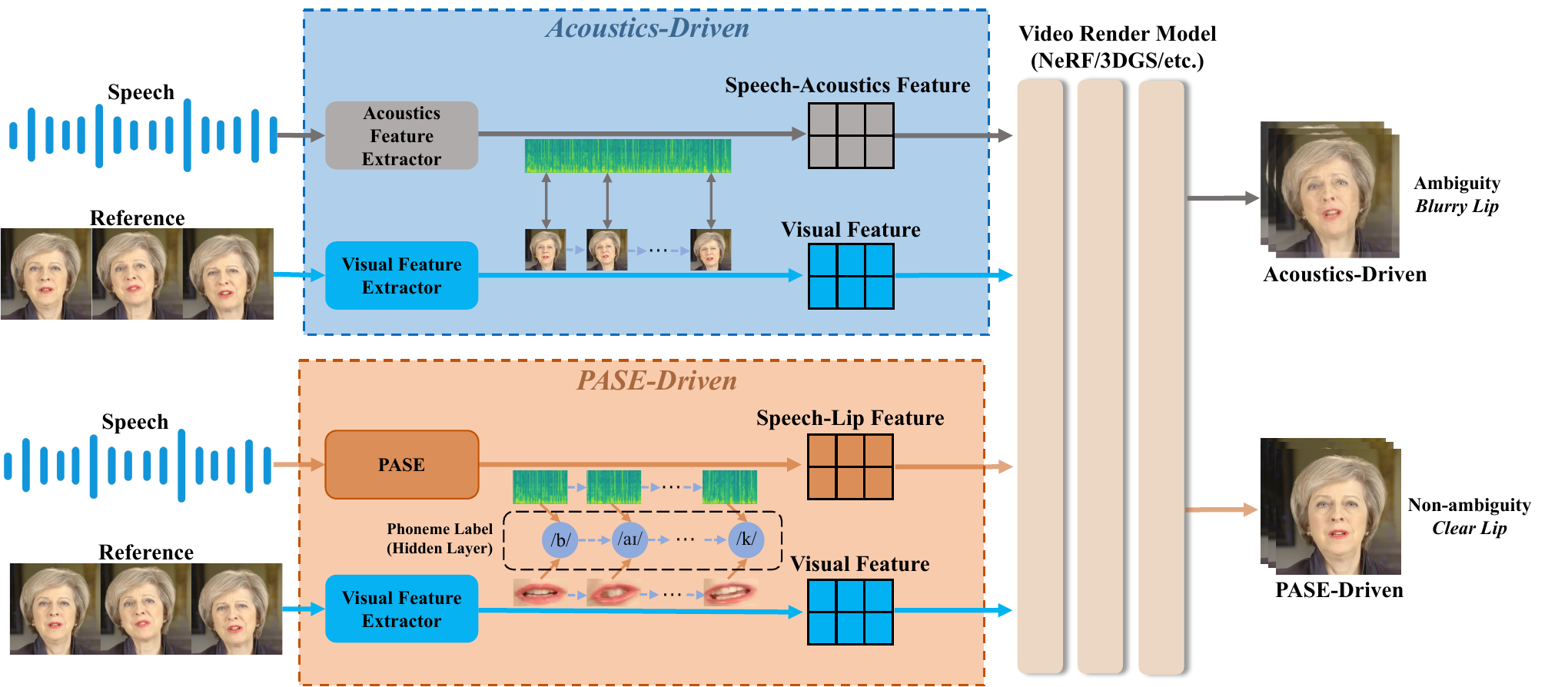}
  \caption{The comparison of the talking head synthesis pipeline when using acoustic features and PASE. The core of PASE is to solve phoneme-viseme alignment ambiguity, which refers to the uncertainty and imprecision in matching phonemes (speech) with visemes (lip). PASE is an independent encoder that can be seamlessly integrated into various rendering models to enhance the quality of synthesized faces.}
  \label{application}
\end{figure*}

\IEEEPARstart{T}{alking } head synthesis has attracted widespread applications in video conferences \cite{video_conference}, film production \cite{film_making}, psychology \cite{psychological}, and other fields. There is an expectation to generate dynamic, realistic, and stable synthetic videos, especially regarding lip movements. The pipeline of talking head synthesis is illustrated in Figure \ref{application}. Speech and visual features are used as conditional inputs to a rendering model, which ultimately synthesizes the video. Since this task is driven by speech, the quality of the synthesized video heavily depends on the quality of the speech features. 

Existing work uses general acoustic features as guided speech features. However, we discovered that these features suffer from phoneme-viseme alignment ambiguity. Ambiguity refers to the phenomenon where different phonemes (e.g., /t/ and /d/, /k/ and /\textipa{g}/) correspond to similar visemes (lip). As shown in Figure \ref{application}, we first decompose the phoneme–viseme alignment task into two generalizable sub-tasks: 1) extracting visual motion information from acoustic features; 2) mapping the extracted motion information to visual image sequences. As shown by the upper part in Figure \ref{application}, current work employs acoustic features, such as DeepSpeech \cite{deepspeech}, HuBERT \cite{hubert}, Wav2Vec 2.0 \cite{wav2vec} and Whisper \cite{whisper}, as conditional input to the rendering model for video synthesis. However, these features are designed for tasks like speech recognition or speaker identification and focus on the discriminative power of phonemes in the acoustic domain. Consequently, phoneme-viseme alignment relies on weak alignment within the rendering model, which is simply achieved through methods like feature addition or attention mechanisms. This weak alignment does not adequately address the ambiguity between phonemes and visemes. As a result, several issues can arise when using acoustic features to drive video synthesis: \textbf{1) Inaccurate lip shape.} Due to the phoneme-viseme alignment ambiguity, the alignment between the lip movements and the phonemes is inaccurate. \textbf{2) Blurry lip shape.} Acoustic features do not learn the dynamic relationship between the speech signal and lip movements, resulting in a blurry lip shape. 

This paper proposes a universal speech encoder, PASE, which can be integrated into various talking head synthesis models to solve the issue of phoneme-viseme alignment ambiguity. PASE introduces a phoneme-aware multimodal alignment framework with contrastive learning. Specifically, PASE models the audio using an STFT spectrogram and a GRU-based encoder, which preserves high-frequency phonetic cues and captures temporal dynamics at the phoneme level. On the visual side, inspired by Wav2Lip \cite{wav2lip} and SyncNet \cite{syncnet}, PASE employs a multi-channel CNN to extract spatio-temporal lip features. To enhance robustness, a prediction–reconstruction task is applied, where portions of audio or visual features are randomly masked and reconstructed during training. Finally, PASE introduces a phoneme-level alignment module that explicitly enforces the audio–visual correspondence. For each phoneme, audio features serve as anchors and are fused with visual features via a cross-attention mechanism. To further enhance phoneme discrimination, phoneme embeddings are integrated into the audio features before fusion. Contrastive learning is then applied, minimizing the distance between positive audio–visual pairs while maximizing the distance to negative pairs. This design ensures that PASE focuses on phoneme-specific articulatory patterns. As illustrated in the lower part of Fig. \ref{application}, PASE leverages the semantic layer (phonemes) shared between the speech and visual modalities as alignment anchors. This design exploits not only temporal relationship constraints but also the stable semantic structure, thereby enhancing the consistency of cross-modal alignment. Experimental results show that videos synthesized using PASE outperform those synthesized using acoustic features in both NeRF and 3DGS rendering models. We summarize our contributions as follows.

\begin{enumerate}
    \item [1)] \textbf{Universal Speech-Lip Encoder}:
    To address the phoneme-viseme alignment ambiguity in talking head synthesis, we propose a universal speech-lip encoder that can be seamlessly integrated into various synthesis models. This encoder directly establishes alignment between speech and lip features within a shared feature space, enabling the extracted speech features to carry rich visual information.

    \item [2)] \textbf{Phoneme-Level Alignment Framework:} We propose a phoneme-level alignment framework that integrates phoneme embeddings into the audio features and employs cross-attention with visual features. By optimizing with contrastive learning at the phoneme granularity, this framework explicitly enforces discriminative correspondence between phonemes and visemes, significantly reducing alignment ambiguity.

    \item [3)] \textbf{Excellent Performance:} Compared to four acoustic features, we achieve state-of-the-art results on both NeRF and 3DGS rendering models. Our method achieves a 13.7\% improvement in lip sync error confidence and a 14.2\% improvement in lip sync error distance compared to the best baseline. The results are also closely aligned with ground truth videos. Furthermore, the results of the ablation experiments validate the effectiveness of the STFT spectrogram and the GRU-based model.
\end{enumerate}

\section{Related Work}
\subsection{Acoustic Feature}
Acoustic features aim to extract linguistic representations from speech signals for subsequent tasks such as speech recognition and speaker identification. Representative works include DeepSpeech \cite{deepspeech}, HuBERT \cite{hubert}, Wav2Vec 2.0 \cite{wav2vec} and Whisper \cite{whisper}. Although these features perform excellently in downstream tasks such as speech recognition and speaker identification, they are not designed with the temporal and frequency detail requirements of the talking head synthesis task in mind. They are designed to extract linguistic representations from the speech signal but cannot enhance the rendering model's alignment capability. 

\subsection{Talking Head Synthesis}
In recent years, due to its applications in digital humans, virtual avatars, and video conferencing, talking head synthesis \cite{gan_1,gan_2,gan_3,gan_4,gan_5,ad_nerf,er_nerf,synctalk,geneface,gaussianspeech,talkinggaussian,loopy,difftalk,diffusion_1,diffusion_2,diffusion_3}, especially the real-time talking head synthesis \cite{rad_nerf,ad_nerf,ae_nerf,synctalk,geneface,geneface++,sd_nerf,er_nerf++,gaussian_flow,gaussiantalker,emotalk3d,gaussianspeech,talkinggaussian}, has attracted significant attention. The main rendering models currently used are based on Neural Radiance Fields (NeRF) \cite{nerf} and 3D Gaussian Splatting (3DGS) \cite{3d_gaussian}. NeRF achieves high-fidelity scene rendering by constructing an implicit continuous volume scene representation and modeling the color and density distribution of light propagation using multi-layer perceptions. 3D Gaussian Splatting, on the other hand, parameterizes the scene geometry and appearance using dynamic Gaussian point clouds and realizes real-time dynamic rendering through differentiable rasterization. In some representative works, AD-NeRF \cite{ad_nerf}, ER-NeRF \cite{er_nerf} and TalkingGaussian \cite{talkinggaussian} use DeepSpeech as the speech feature; GeneFace \cite{geneface} uses HuBERT; GaussianSpeech \cite{gaussianspeech} uses Wav2Vec 2.0; and Salehi et al. \cite{comparative} use Whisper. 
\begin{table*}[tb]
  \centering
    \caption{The list of some phonemes (using consonants as an example) related to phoneme-viseme alignment ambiguity. “\&” indicates that the lip shapes of the two phonemes are similar.}
    \scalebox{0.95}{\begin{tabular}{ccc}
    \toprule
    \textbf{Classification} & \textbf{Manner of Pronunciation} & \textbf{Examples} \\
    \midrule
         Plosive / Stop & Complete closure in the mouth and sudden release of lung air through the mouth & /p/\&/b/, /t/\&/d/, /k/\&/\textipa{g}/ \\
         Nasal & Complete oral closure in the mouth, the air escapes through the nose & /m/\&/n/\&/\textipa{N}/ \\
         Fricative & Narrowing with audible friction, close approximation & /f/\&/v/, /s/\&/z/, /\textipa{T}/\&/\textipa{D}/, /\textipa{S}/\&/\textipa{Z}/ \\
         Affricate & Complete oral closure and slow release of the lung air & /\textteshlig/ \& /\textdyoghlig/ \\
    \bottomrule
    \end{tabular}%
    }

  \label{phoneme}
\end{table*}%

Acoustic features are optimized for phoneme classification, focusing on the acoustic distinction between phonemes rather than the alignment between phonemes and visemes. In the pipeline of traditional methods, phoneme-viseme alignment relies solely on the rendering model. However, existing rendering models often align phonemes and visemes through simple feature addition or attention mechanisms, which can not avoid the phoneme-viseme alignment ambiguity.

\section{Motivation}

\begin{figure}[t]
    \centering
    \includegraphics[width=\linewidth]{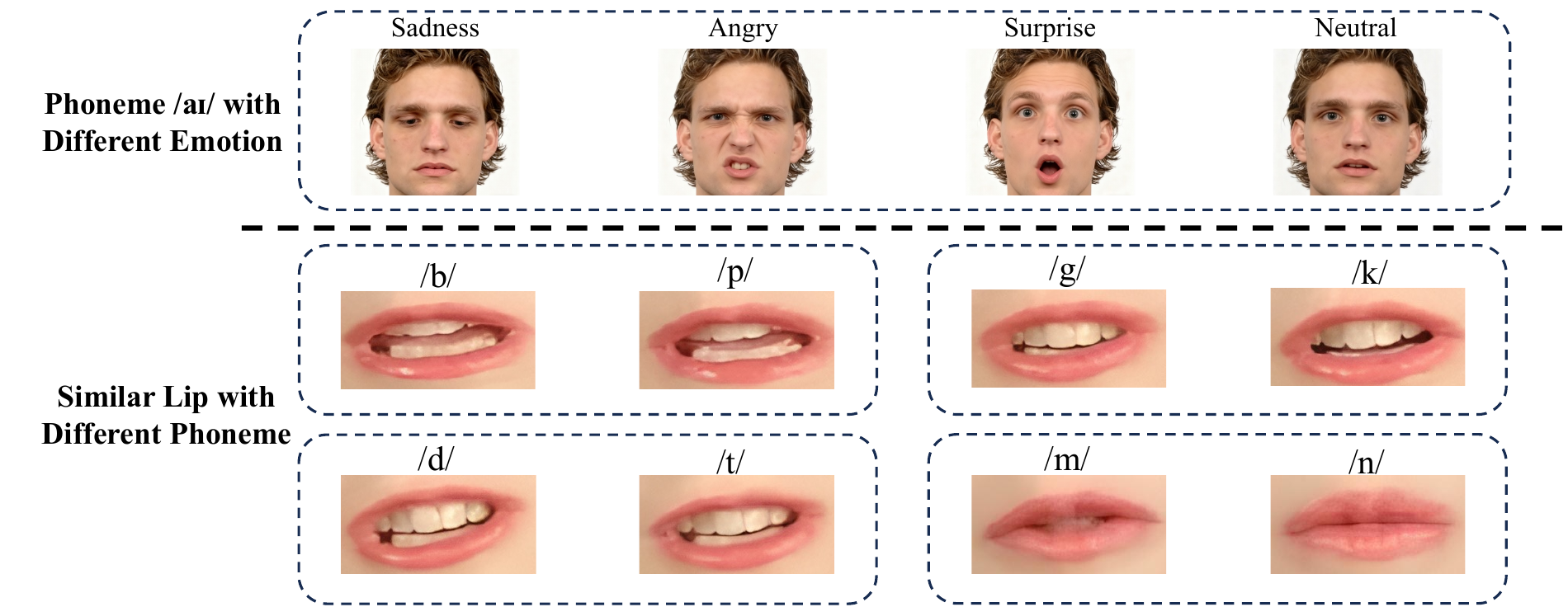}
    \caption{The illustration of the many-to-many relationship between speech and visual modalities. Similar lip shapes correspond to different phonemes, while the same phoneme corresponds to different lip shapes under different emotions.}
    \label{lips}
\end{figure}

\begin{figure}[t]
    \centering
    \includegraphics[width=0.8\linewidth]{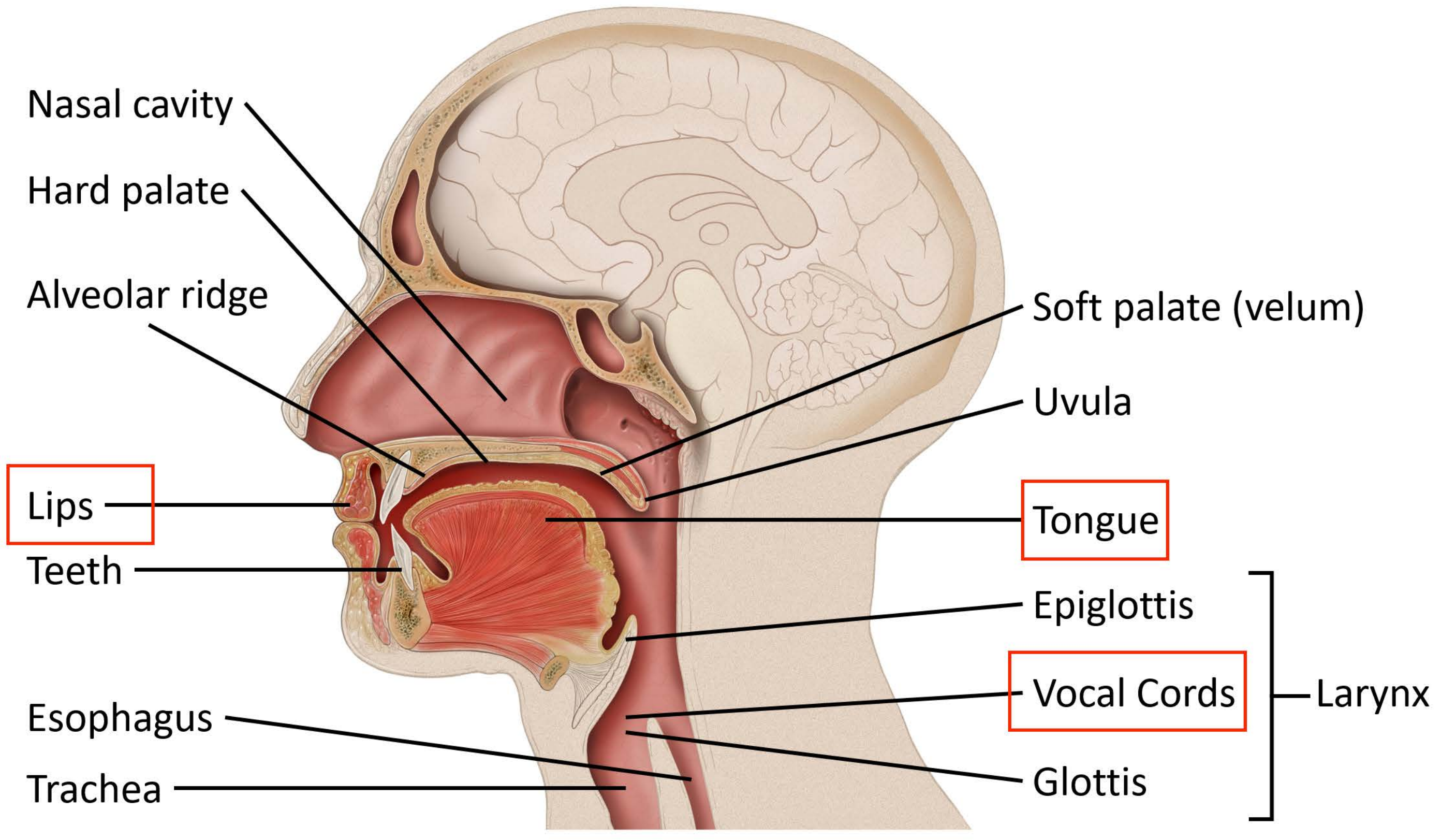}
    \caption{The vocal tract anatomy diagram. The lips are only one of the organs that affect pronunciation.}
    \label{vocal}
\end{figure}


\begin{figure}[tb]
\centering
    {\includegraphics[width=\linewidth]{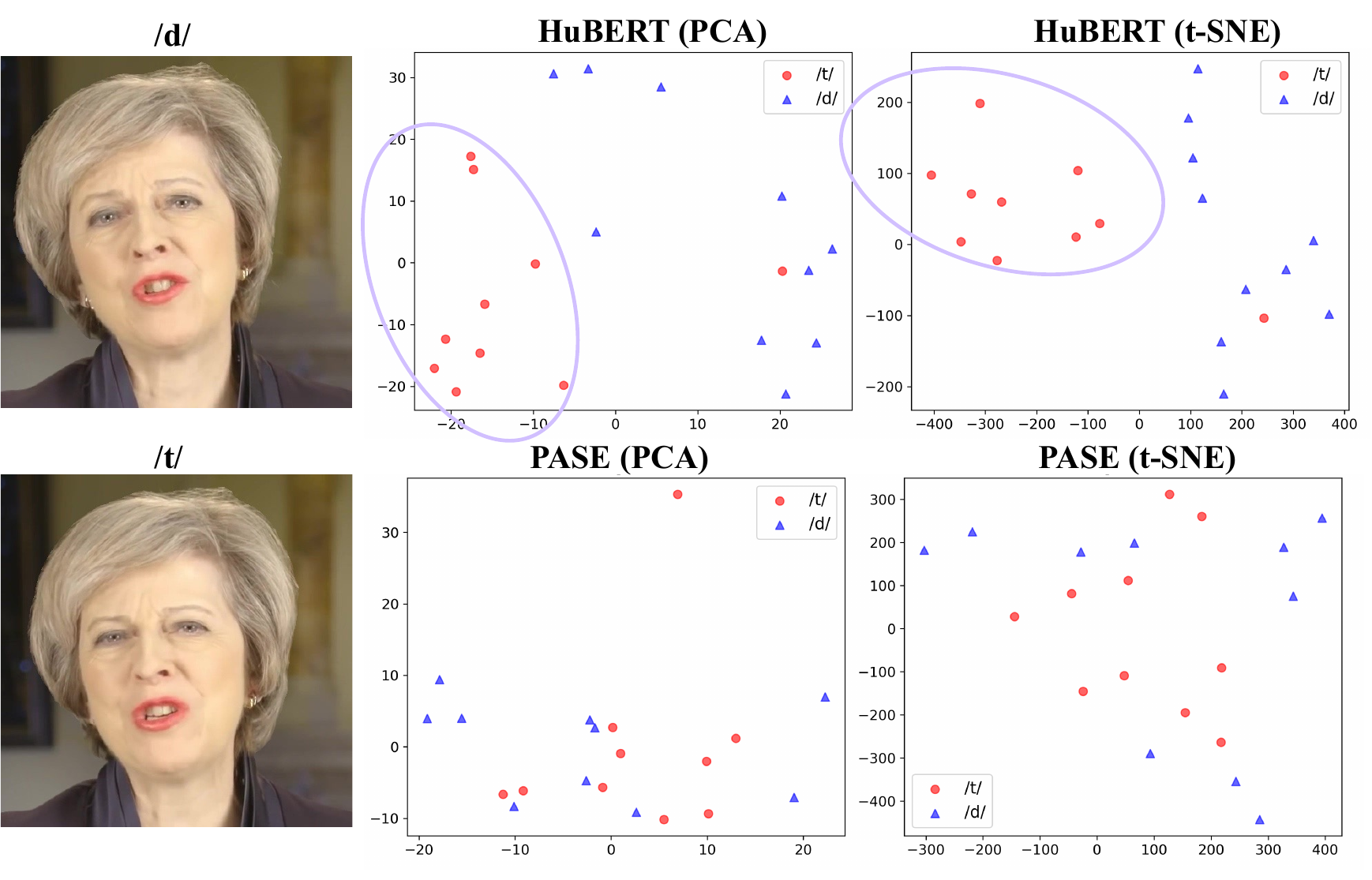}}
    \caption{Taking the phonemes /d/ and /t/ as examples to illustrate the phoneme-viseme alignment ambiguity. In the visualization of HuBERT \cite{hubert} features, there is a noticeable difference between /d/ and /t/. In the visualization of PASE features, there is no significant distinction between /d/ and /t/ because they share similar lip shapes, indicating that PASE differentiates phonemes based on lip shapes rather than acoustic features.}
    \label{motivation}
\end{figure}

\subsection{Phoneme-Viseme Alignment Ambiguity}
As shown in Figure \ref{lips}, the relationship between speech and visual modalities is inherently many-to-many. For certain types of phonemes (e.g., plosives), similar lip shapes may correspond to different phonemes. Moreover, the same phoneme can produce inconsistent lip shapes under different emotional states. This many-to-many relationship results in phoneme–viseme alignment ambiguity. In Table \ref{phoneme}, we list some phonemes related to phoneme-viseme alignment ambiguity along with their corresponding pronunciation manner. To explain the origin of this phenomenon in detail, we also present a vocal tract anatomical diagram in Figure \ref{vocal} from Ramoo \cite{ramooPsychologyLanguage2021} to illustrate the organs that affect pronunciation. The larynx or vocal cords are the basis of pronunciation, while the lips and tongue are the articulatory organs. Although the lips are only a part of the speech production system, they are the most intuitive visual feature, which leads to the phoneme-viseme alignment ambiguity issue. For example, as shown in Figure \ref{motivation}, the lip shapes corresponding to phonemes /d/ and /t/ are similar. However, /d/ is a voiced consonant with strong vocal cord vibrations during pronunciation, while /t/ is a consonant with weak vocal cord vibrations. Despite the significant acoustic differences between /d/ and /t/, the speech features of /d/ and /t/ should be aligned to similar visual features in the cross-modal alignment.

\subsection{Drawback of Using Acoustic Features}

Acoustic-driven talking head synthesis suffers from the issue of phoneme-viseme alignment ambiguity. The root cause lies in two drawbacks:
1) \textbf{Misalignment of Task Objectives.} Acoustic features aim to extract highly discriminative linguistic representations, which directly conflict with the synthesis goal of aligning phonemes and lip shapes. As shown in the scatter plot of Figure \ref{motivation}, we extracted HuBERT \cite{hubert} features for /d/ and /t/ and visualized them using PCA \cite{pca} and t-SNE \cite{t-sne} methods. It is evident that due to the difference in linguistic representations, HuBERT strongly distinguishes between the two phonemes. However, the talking head synthesis task requires alignment between the phoneme and visual features (particularly the lip shape). This mismatch forces rendering models to rely on weak alignment mechanisms like feature addition or attention, exacerbating ambiguity and reducing lip-shape accuracy. \textbf{2) Lack of Dynamic Representation.} The temporal resolution of acoustic features (typically 25ms frame length) struggles to capture the instantaneous changes in lip movements. For example, the lip closure-opening process for the plosive sound /p/ lasts around 80–120ms. However, the dynamic features of this movement are dropped by the Mel spectrogram in the high frequency ranges, leading to a blurry lip shape in the synthesis.

\subsection{Contrastive Learning of Speech-Lip}
To enhance the alignment capability of the rendering model and address the issue of phoneme-viseme alignment ambiguity, we propose using cross-modal speech features as input for the rendering model. To train such a speech encoder, we use contrastive learning and directly establish the alignment between speech and lip features in a joint embedding space. Our approach forces the speech encoder to focus on the causal relationship between phonemes and visemes rather than speech content or speaker identity. Additionally, we preserve fine-grained speech feature information through detailed processing in both the frequency and time domains. Specifically, we propose using the STFT spectrogram instead of the Mel spectrogram as input to the speech encoder to avoid the loss of high frequency ranges. We also introduce a temporal model to improve the temporal resolution of speech features.

\section{Methodology}\label{method}

\subsection{Overview}
\begin{figure*}[t]
\centering
    {\includegraphics[width=1.0\linewidth]{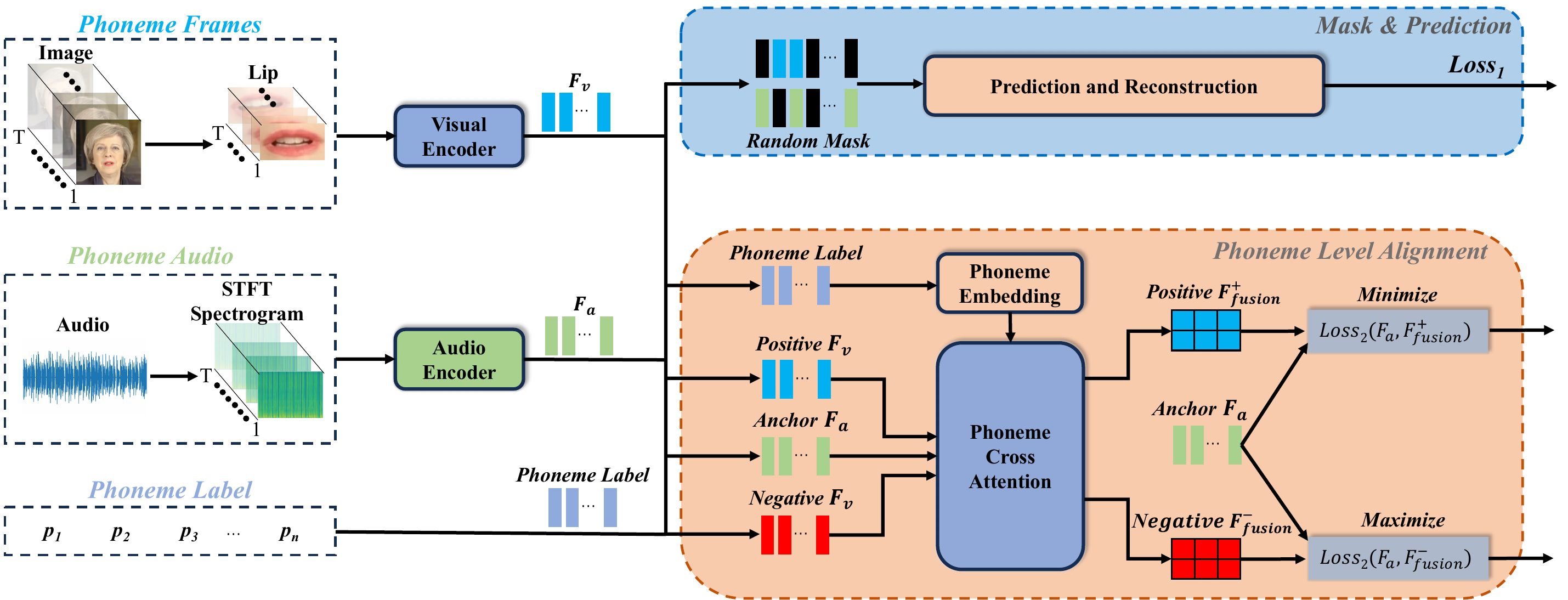}}
    \caption{The framework of PASE.}
    \label{framework}
\end{figure*}

The framework of PASE is illustrated in Figure \ref{framework}, which consists of the Audio Encoder, the Visual Encoder, the Prediction and Reconstruction module, and the Phoneme Level Alignment module. During the data preprocessing stage, we first collect the raw images and speech segments corresponding to the same phoneme interval. We extract lip images from the video on a frame-by-frame level, and convert the corresponding speech segments into STFT spectrograms. In the Audio Encoder, we employ a gated recurrent unit (GRU) as the temporal model to extract audio features, thereby capturing critical sequential information from audio. In the Visual Encoder, we design a multi-channel convolutional neural network (CNN) to extract lip-related visual features. The multi-channel CNN is capable of simultaneously capturing both temporal variations and spatial characteristics of lip movements. To enhance robustness, we apply the Prediction and Reconstruction module to both audio and visual features. This module randomly masks a portion of the audio or visual features and requires the model to predict and reconstruct the masked parts using the remaining unmasked features. In the Phoneme Level Alignment module, we treat the audio features as anchors and perform feature fusion through a cross-attention mechanism. During feature fusion, phoneme embeddings are incorporated to enhance the model’s ability to recognize and discriminate specific phonemes. Finally, a contrastive loss is employed to minimize the distance between positive pairs while maximizing the distance between negative pairs, thereby achieving phoneme-level multimodal feature alignment.

\subsection{Audio Encoder}
Most existing methods follow the traditional paradigm by directly using the Mel spectrogram, without thoroughly investigating their suitability for phoneme-viseme alignment tasks. We argue that the STFT spectrogram is a more appropriate choice. The Mel spectrogram is originally designed to mimic human auditory perception by preserving formant structures (typically within 0.3–4kHz) that convey semantic information, while suppressing high-frequency components to improve efficiency and robustness in speech recognition. However, high-frequency components (>4kHz) carry important articulatory motion cues. For instance, fricatives like /\textipa{S}/ and /s/ exhibit distinct energy distributions in the high-frequency. However, the high-frequency suppression in the Mel spectrogram undermines the model’s ability to capture dynamic visual cues, thereby increasing the alignment ambiguity. Furthermore, the superiority of STFT is also validated through the ablation studies \ref{ablation study}.

The extracted STFT spectrogram is processed through a GRU network. The core of GRU consists of the update gate $z_t$ and the reset gate $r_t$. For the STFT spectrogram input sequence$\{x_1,...,x_T\}$, the definitions of $z_t$ and $r_t$ are given in Eq. \ref{gru_gates}.

\begin{equation}
\begin{aligned}
    z_t &= \sigma(W_z \cdot [h_{t-1}, x_t]) \\
    r_t &= \sigma(W_r \cdot [h_{t-1}, x_t])
\end{aligned}
\label{gru_gates}
\end{equation}

Here, $\sigma$ denotes the Sigmoid function, $h_{t-1}$ is the hidden state at the previous time step, and $W_z$ and $W_r$ are the corresponding weight matrices. The computation process of GRU at time step $t$ is represented by Eq. \ref{hidden}. 

\begin{equation}
\begin{aligned}
        \tilde{h}_t &= \tanh(W_h \cdot [r_t \odot h_{t-1}, x_t]) \\
    h_t &= (1-z_t) \odot h_{t-1} + z_t \odot \tilde{h}_t 
\end{aligned}
\label{hidden}
\end{equation}

Here, $\tilde{h}_t$ is the candidate hidden state, and $h_t$ is the final hidden state. We select the output of the last GRU layer as the final feature. This network effectively models the temporal feature of speech. It captures the long-term dynamic changes in speech. This is particularly important in the talking head synthesis task, where the temporal relationships between phonemes are crucial. Therefore, we choose GRU instead of the commonly used CNN to better model these temporal dependencies. Through the output of the GRU, we obtain a speech embedding that contains dynamic information, providing strong support for the subsequent alignment of the rendering model. 

\subsection{Visual Encoder}
\begin{table}[htbp]
    \caption{The network architecture of the lip feature extractor.}
  \centering
    \scalebox{0.75}{\begin{tabular}{cccccc}
    \toprule
    \textbf{Layer Type} & \textbf{Input Chanel} & \textbf{Output Chanel} & \textbf{Kernel Size} & \textbf{Stride} & \textbf{Padding} \\
    \midrule
    \textbf{Conv2d} & 15    & 32    & 7×7   & 1     & 3 \\
    \textbf{Conv2d} & 32    & 64    & 5×5   & (1,2) & 1 \\
    \textbf{ResidualBlock	} & 64    & 64    & 3×3   & 1     & 1 \\
    \textbf{ResidualBlock	} & 64    & 64    & 3×3   & 1     & 1 \\
    \textbf{Conv2d} & 64    & 128   & 3×3   & 2     & 1 \\
    \textbf{ResidualBlock	} & 128   & 128   & 3×3   & 1     & 1 \\
    \textbf{ResidualBlock	} & 128   & 128   & 3×3   & 1     & 1 \\
    \textbf{Conv2d} & 128   & 256   & 3×3   & 2     & 1 \\
    \textbf{ResidualBlock	} & 256   & 256   & 3×3   & 1     & 1 \\
    \textbf{Conv2d} & 256   & 512   & 3×3   & 2     & 1 \\
    \textbf{ResidualBlock	} & 512   & 512   & 3×3   & 1     & 1 \\
    \textbf{Conv2d} & 512   & 512   & 3×3   & 2     & 1 \\
    \textbf{Conv2d} & 512   & 512   & 3×3   & (4,1) & 0 \\
    \textbf{Conv2d} & 512   & 512   & 1×1   & 1     & 0 \\
    \bottomrule
    \end{tabular}%
    }
  \label{network_1}
\end{table}%
We aim to provide accurate lip features for aligning speech and lip shapes. Inspired by works \cite{wav2lip,syncnet}, we designed a multi-channel CNN-based model to extract lip features. The specific architecture of this model can be found in Table \ref{network_1}. The model captures fine-grained spatial features of the lip images layer by layer through multiple convolutional layers. After processing the lip shape image through the CNN, an embedding is obtained with the same dimension as the speech features, representing the dynamic visual features of the lip shapes.

\subsection{Prediction and Reconstruction}
To enhance the robustness of the model against feature noise, we introduce a Prediction and Reconstruction task. For the visual features $V \in \mathbb{R}^{T \times D}$ and audio feature sequence $A \in \mathbb{R}^{T \times D}$ belonging to the same phoneme segment, we generate separate binary masking matrices. These matrices randomly select a subset of time steps, where the corresponding features are replaced with learnable vectors. The masking operations for the visual sequence $V$ and the audio sequence $A$ are performed independently and randomly, meaning that one modality, or both simultaneously, can be masked. During training, the model is required to predict and reconstruct the masked segments, thereby providing additional supervisory signals for cross-modal representation learning. The reconstruction loss is defined as Eq. \ref{rec_loss}.
\begin{equation}\label{rec_loss}
    \mathcal{L}_{rec} = \lvert\lvert V_{masked} -V_{orig} \rvert\rvert_2^2 + \lvert\lvert A_{masked} - A_{orig} \rvert\rvert_2^2
\end{equation}
This auxiliary loss branch effectively alleviates overfitting and enhances the model’s robustness when facing occlusions or modality absence.

\subsection{Phoneme Level Alignment}
To improve the alignment capability of the model, we propose a phoneme-level alignment strategy. Specifically, for a given phoneme $p$, we use the corresponding audio feature $A_p$ as the anchor and apply a cross-attention mechanism to fuse positive and negative visual features. During the fusion process, the phoneme label is embedded to obtain the phoneme label feature $F_p$. As formulated in Eq. \ref{q_vector}, the phoneme label feature is integrated with the audio feature to generate the query vector $Q_p$.
\begin{equation}
\label{q_vector}
    Q_p = A_p + F_p
\end{equation}
This design enhances the model’s phoneme-awareness during feature fusion. Subsequently, the visual features $V_p$ are used as both key and value vectors, and the fused feature $F_{fusion}$ is obtained according to Eq. \ref{cross_attn}.
\begin{equation}\label{cross_attn}
    F_{fusion} = CrossAttn(Q_p,V_p,V_p)
\end{equation}
For the positive pair $(A_p, F^+_{fusion})$ and negative pairs $(A_p, F^-_{fusion})$, we employ a contrastive loss, as defined in Eq. \ref{cl_loss}, to constrain the alignment. 

\begin{equation}\label{cl_loss}
\resizebox{0.9\columnwidth}{!}{
    $\mathcal{L}_{\text{con}} = - \log \frac{\exp \left( s(A_p, F^+_{fusion}) / \tau \right)}
{\exp \left( s(A_p, F^+_{fusion}) / \tau \right) + \sum\limits_{F^-_{fusion}} \exp \left( s(A_p, F^-_{fusion}) / \tau \right)}$
}
\end{equation}

Here, $s(\cdot)$ denotes the similarity function, and $\tau$ represents the temperature coefficient. Finally, the overall objective function is defined in Eq. \ref{total_loss}.
\begin{equation}\label{total_loss}
    \mathcal{L} =\mathcal{L}_{\text{con}} + \alpha \cdot \mathcal{L}_{rec}
\end{equation}
In this formulation, $\alpha$ is a weighting factor that balances the reconstruction loss.

\section{Experiments}
\begin{table*}[t]
  \centering
    \caption{The quantitative results of video synthesis using different speech features. We highlight the \textbf{best} and \underline{second best} results.}
    \scalebox{0.9}{\begin{tabular}{r|c|cc|ccc}
    \toprule
          &       & \textbf{PSNR↑(vs. HuBERT)} & \textbf{LPIPS↓(vs. HuBERT)} & \textbf{LMD↓(vs. HuBERT)} & \textbf{LSE-C↑(vs. HuBERT)} & \textbf{LSE-D↓(vs. HuBERT)} \\
    \midrule
          & \textbf{Ground Truth} & N/A   & 0     & 0     & 8.8302 (+15.3\%) & 6.0570 (-15.2\%) \\
    \midrule
          & \textbf{HuBERT \cite{hubert}} & 32.0108 (+0\%) & 0.0427 (+0\%) & 2.9616 (+0\%) & 7.6599 (+0\%) & 7.1402 (+0\%) \\
          & \textbf{DeepSpeech \cite{deepspeech}} & 32.0338 (+0.07\%) & 0.0417 (-2.34\%) & 3.0321 (+2.38\%) & 7.4268 (-3.04\%) & 7.5418 (+5.63\%) \\
     & \textbf{Wav2vec 2.0 \cite{wav2vec}} & 31.6359 (-1.17\%) & 0.0422 (-1.17\%) & 3.7204 (+25.6\%) & 3.3782 (-55.9\%) & 10.200 (+42.8\%) \\
\multicolumn{1}{c|}{\textbf{NeRF}} & \textbf{Whisper \cite{whisper}} & 31.2311 (-2.44\%) & 0.0436 (+2.11\%) & 3.6727 (+24.0\%) & 3.3416 (-56.4\%) & 10.202 (+42.8\%) \\
          & \textbf{AVHubert \cite{shi2022avhubert}} & 31.8361 (-0.54\%) & 0.0421 (-1.41\%) & 2.9273 (-1.16\%) & 8.0136 (+4.62\%) & 6.8022 (-4.73\%) \\
              & \textbf{Wav2Lip \cite{wav2lip}} & 32.1391 (+0.40\%) & \underline{0.0406 (-4.92\%)} & 2.9771 (+0.52\%) & 8.4461 (+10.3\%) & \underline{6.4028 (-10.3\%)} \\
          & \textbf{SyncTalk \cite{synctalk}} & \textbf{32.5362(1.33\%)} & 0.0470(-10.1\%) & \underline{2.8842(-2.61\%)} & \underline{8.6977(11.9\%)} & 6.5323(-8.51\%) \\
          
          & \textbf{PASE(Ours)} & \underline{32.2301 (+0.68\%)} & \textbf{0.0399 (-6.56\%)} & \textbf{2.8725 (-3.01\%)} & \textbf{8.7098 (+13.7\%)} & \textbf{6.1255 (-14.2\%)} \\
    \midrule
          & \textbf{Ground Truth} & N/A   & 0     & 0     & 8.8302 (+74.3\%) & 6.0570 (-34.2\%) \\
    \midrule
          & \textbf{HuBERT \cite{hubert}} & \underline{30.9339 (+0\%)} & 0.0411 (+0\%) & 2.9634 (+0\%) & 5.0660 (+0\%) & 9.2126 (+0\%) \\
          & \textbf{DeepSpeech \cite{deepspeech}} & 30.8336 (-0.32\%) & 0.0415 (+0.97\%) & 3.0547 (+3.08\%) & 4.7966 (-5.31\%) & 9.4243 (+2.29\%) \\
     & \textbf{Wav2vec 2.0 \cite{wav2vec}} & 30.8003 (-0.42\%) & 0.0421 (+2.43\%) & 3.2513 (+9.72\%) & 3.9522 (-21.9\%) & 9.6083 (+4.29\%) \\
          \multicolumn{1}{c|}{\textbf{3DGS}}& \textbf{Whisper \cite{hubert}} & 30.5449 (-0.28\%) & 0.0410 (-0.24\%) & 3.1397 (+5.93\%) & 3.2848 (-35.1\%) & 10.344 (+12.3\%) \\
          & \textbf{AVHubert \cite{shi2022avhubert}} & 30.8315 (-0.33\%) & \underline{0.0408 (-0.73\%)} & 2.8779 (-2.89\%) & 5.3618 (+5.84\%) & 8.7027 (-5.53\%) \\
          & \textbf{Wav2Lip \cite{wav2lip}} & 30.6718 (-0.85\%) & 0.0415 (+0.97\%) & \underline{2.7758 (-6.33\%)} & \underline{6.9973 (+38.1\%)} & \underline{7.7244 (-16.2\%)} \\
          & \textbf{TalkingGaussian \cite{talkinggaussian}} & 30.8790(-0.17\%) & 0.0411(+0\%) & 3.0011(+1.27\%) & 4.9713(-1.86\%) & 9.0187(-2.1\%) \\
          
          & \textbf{PASE(Ours)} & \textbf{31.0176 (+0.29\%)} & \textbf{0.0401 (-2.43\%)} & \textbf{2.6836 (-9.44\%)} & \textbf{8.3487 (+64.8\%)} & \textbf{6.6705 (-27.6\%)} \\
    \bottomrule
    \end{tabular}%
    }
  \label{quantitative}%
\end{table*}%

\subsection{Experimental Settings}
\noindent\textbf{Dataset.} We use the LRS2 \cite{LRS2} dataset to train PASE.  We follow the current works \cite{synctalk, talkinggaussian, ad_nerf,er_nerf} and use the same well-edited video dataset for the face synthesis. This dataset has a frame rate of 25 fps, a resolution of 512x512, and the subject is centered in the video. 

\noindent\textbf{Comparison Method.} We selected the latest representative works, SyncTalk \cite{synctalk} and TalkingGaussian \cite{talkinggaussian}, as rendering models. Specifically, SyncTalk refers to the NeRF-based framework, while TalkingGaussian refers to the 3DGS-based framework. We further replace the speech encoders in both models with DeepSpeech \cite{deepspeech}, HuBERT \cite{hubert}, Wav2Vec2.0 \cite{wav2vec}, Whisper \cite{whisper}, AVHubert \cite{shi2022avhubert}, Wav2Lip \cite{wav2lip}, and PASE for comparison.

\noindent\textbf{Preprocessing.} We train PASE using the LRS2 \cite{LRS2} dataset. Before training, the dataset undergoes a series of preprocessing steps. To obtain accurate phoneme annotations, we employ the Montreal Forced Aligner \cite{mfa} to align all audio samples in the dataset with corresponding phonemes and timestamps. Based on the alignment results, each video is segmented into phoneme-level audio clips and their associated facial frames. From each audio segment, STFT spectrograms are extracted to serve as the input audio representation. For lip images, we first apply Face Alignment \cite{bulat2017far} to detect facial landmarks. We then retain only the coordinates corresponding to the lip region (landmarks 48 to 68) to define the lip boundaries. Finally, the lip region is cropped from the full-face image based on the detected landmarks.

\noindent\textbf{Implementation Details.} PASE uses STFT with hyperparameters is set to n-fft=512, win-length=512, and hop-length=128. The number of GRU layers is set to 8. During training, the learning rate is 5e-5, the window is 5, and the batch size is 16. For SyncTalk, we use SmoothL1 Loss \cite{smoothl1} as the loss function. The training steps are set to 60,000, with the fine-tuning steps set to 88,000. For TalkingGaussian, we use L1 Loss as the loss function, and the training steps are set to 20,000. Both training and video rendering are performed on a single NVIDIA RTX 4090 GPU.

\subsection{Quantitative Evaluation}
\noindent\textbf{Metrics.} For image fidelity, we use the Peak Signal-to-Noise Ratio (PSNR) to measure overall quality and the Learned Perceptual Image Patch Similarity (LPIPS) \cite{lpips} to measure fine details. Additionally, we evaluate the accuracy of lip shapes using the landmark distance (LMD), Lip Sync Error Confidence (LSE-C) \cite{wav2lip}, and Lip Sync Error Distance (LSE-D) \cite{wav2lip}.

\begin{table}[htbp]
    \caption{The quantitative results of lip accuracy. We highlight the \textbf{best} and \underline{second best} results.}
    \centering
\scalebox{0.78}{\begin{tabular}{ccccccc}
    \toprule
          & \multicolumn{3}{c}{\textbf{Audio A}} & \multicolumn{3}{c}{\textbf{Audio B}} \\
    \midrule
          & \textbf{LMD↓} & \textbf{LSE-C↑} & \textbf{LSE-D↓} & \textbf{LMD↓} & \textbf{LSE-C↑} & \textbf{LSE-D↓} \\
    \midrule
    \textbf{HuBERT \cite{hubert}} & 3.0471 & 6.5291 & 7.9572 & 3.0328 & 6.8863 & 7.8993 \\
    \textbf{DeepSpeech \cite{deepspeech}} & 3.1169 & 6.7682 & 7.6824 & 3.1052 & 6.3795 & 8.0862 \\
    \textbf{Wav2Vec 2.0 \cite{wav2vec}} & 3.7019 & 3.8128 & 9.8629 & 3.6718 & 3.5824 & 10.1847 \\
    \textbf{Whisper \cite{whisper}} & 3.7177 & 4.0613 & 9.5127 & 3.6791 & 3.7153 & 10.9471 \\
    \textbf{Synctalk \cite{synctalk}} & \underline{2.9521} & \textbf{8.3814} & \underline{7.2186} & \underline{2.9729} & \underline{7.7299} & \underline{7.3170} \\
    \textbf{PASE(Ours)} & \textbf{2.9304} & \underline{8.1294} & \textbf{6.9714} & \textbf{2.9362} & \textbf{8.0932} & \textbf{7.0046} \\
    \bottomrule
    \end{tabular}%
    }
    \label{diff-samples}
\end{table}

\noindent\textbf{Evaluation Results.} We present the quantitative results in Table \ref{quantitative}. Our method exhibits significant advantages in lip sync accuracy. Our method's LMD surpasses the four acoustic features. In the NeRF model, our method improves LSE-C and LSE-D by 13.7\% and 14.2\%, respectively, compared to the HuBERT \cite{hubert}. This advantage is further amplified in the 3DGS model. Notably, the LSE-C and LSE-D of our method are close to the ground truth video, indicating the fine-grained capture of speech dynamic information by the STFT-GRU combination and the effectiveness of the cross-modal alignment framework. Furthermore, our approach outperforms state-of-the-art face synthesis methods, including AVHubert \cite{shi2022avhubert}, Wav2Lip \cite{wav2lip}, SyncTalk \cite{synctalk}, and TalkingGaussian \cite{talkinggaussian}. It is worth noting that SyncTalk \cite{synctalk} employs specially designed speech features, while TalkingGaussian \cite{talkinggaussian} relies on DeepSpeech \cite{deepspeech}. We also used out-of-distribution speech to drive video synthesis in the NeRF model. We present the experimental results in Table \ref{diff-samples}. We used LMD, LSE-C, and LSE-D as metrics. The experimental results show that our method outperforms the four acoustic features in terms of lip sync accuracy.

\begin{table*}[tb]
  \caption{The results of the user study. The rating is on a scale of 1-5. A higher scores indicate better performance. We highlight the \textbf{best} and \underline{second best} results.}
  \centering
    \scalebox{0.9}{\begin{tabular}{cccccc}
    \toprule
    \textbf{Method} & \textbf{DeepSpeech \cite{deepspeech}} & \textbf{HuBERT \cite{hubert}} & \textbf{Wav2Vec 2.0 \cite{wav2vec}} & \textbf{Whisper \cite{whisper}} & \textbf{PASE(Ours)} \\
    \midrule
    \textbf{Lip-sync Accuracy} & 2.89  & \underline{3.17}  & 2.54  & 2.49  & \textbf{4.27} \\
    \textbf{Image Quality} & 3.82  & 3.78  & \underline{3.89}  & 3.69  & \textbf{4.02} \\
    \textbf{Video Realness} & 3.38  & \underline{3.42}  & 3.37  & 3.19  & \textbf{3.82} \\
    \bottomrule
    \end{tabular}%
    }
  \label{user study}%
\end{table*}%

\subsection{Qualitative Evaluation}
\begin{figure*}[t]
\centering
    {\includegraphics[width=1.0\textwidth]{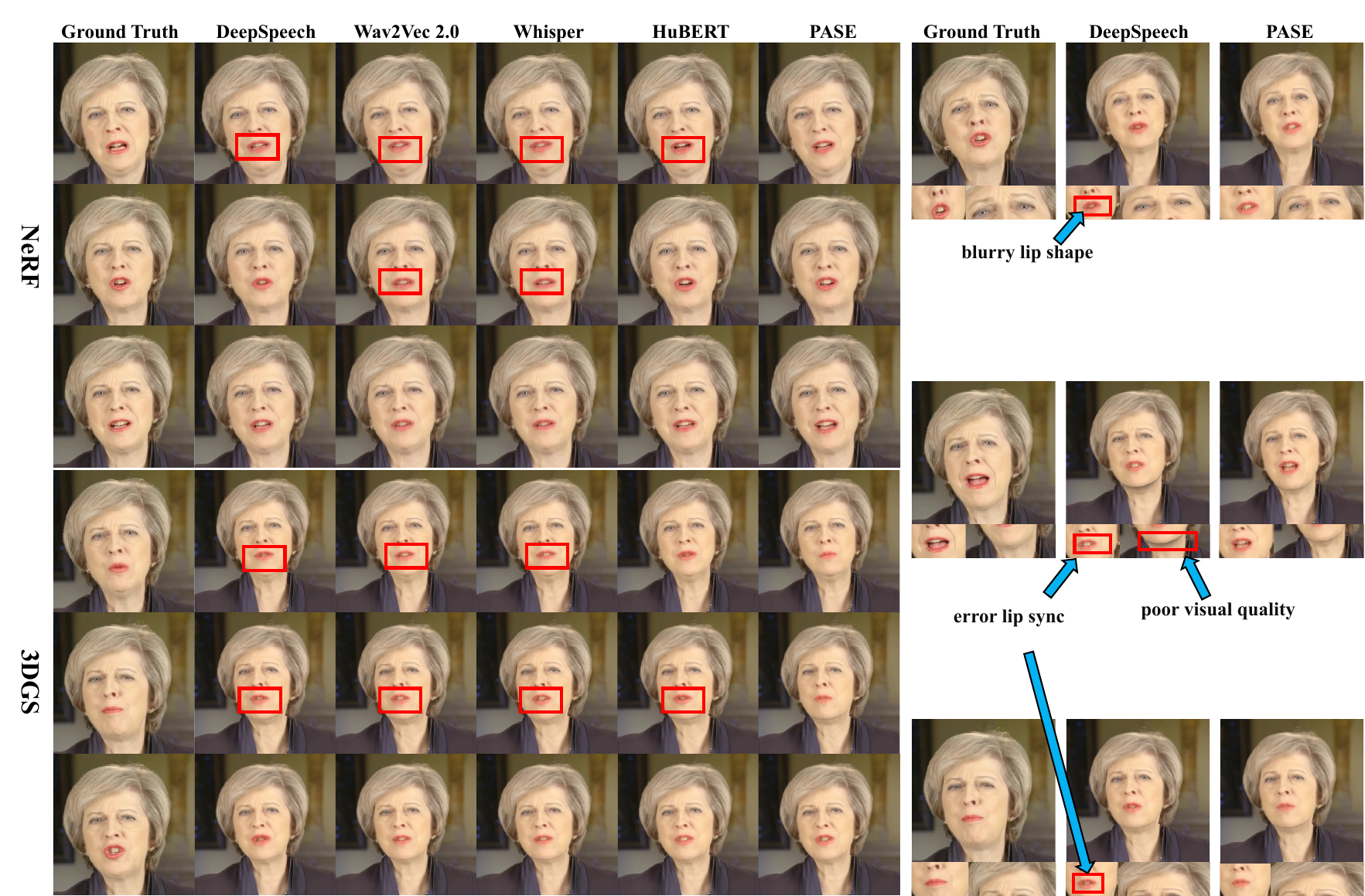}}
    \caption{The qualitative results of video synthesis using different speech features. Our method has the most accurate lip shapes and achieves the best visual quality.}
    \label{qualitative}
\end{figure*}
\noindent\textbf{Evaluation Results.} To intuitively assess the quality of the synthesized video, we present a comparison between our method and other speech features in Figure \ref{qualitative}. As shown, PASE presents more accurate lip shapes and higher visual quality. In the right half of Figure \ref{qualitative}, we provide a detailed comparison between DeepSpeech \cite{deepspeech} and our method. When the movement amplitude of the lip is large, DeepSpeech generates a blurry lip shape (as shown in the first row) and also synthesizes an inaccurate lip shape (as shown in the second row). When the lips are fully closed, DeepSpeech reveals the teeth and fails to fully close the lips (as shown in the third row). Our method can closely approach the ground truth video. This is due to the cross-modal alignment framework, which effectively addresses the phoneme-viseme alignment ambiguity issue. And fine-grained speech feature modeling preserves richer feature details. Furthermore, we observe that inaccurate lip features also affect other parts of the image. For example, in the second row of the right half of the figure, the neck area of the subject shows a noticeable shadow when using DeepSpeech. 

\begin{figure}[tb]
\centering
    {\includegraphics[width=0.93\linewidth]{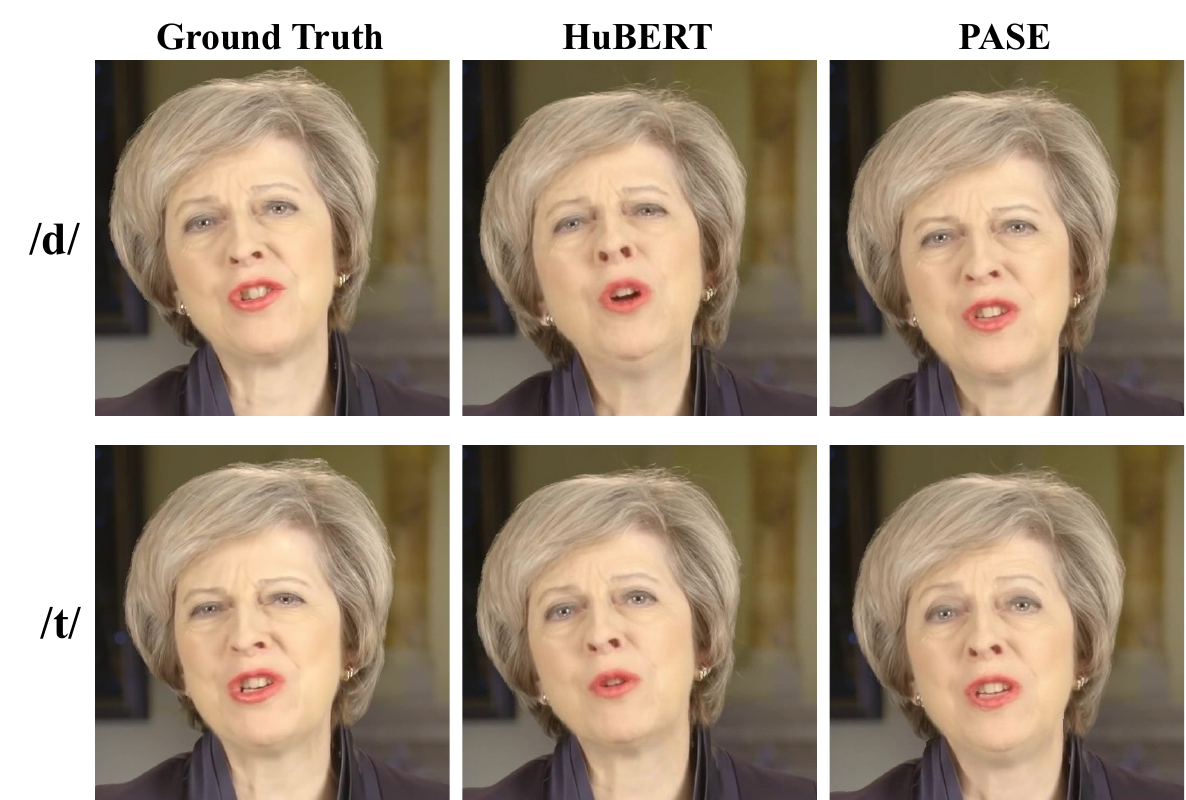}}
    \caption{The qualitative results of phonemes /d/ and /t/. HuBERT \cite{hubert} exhibits the phoneme-viseme alignment ambiguity issue, while our approach avoids this issue.}
    \label{compare_t_d}
\end{figure}

In addition, we also present the synthesis results of phonemes /d/ and /t/ in Fig. \ref{compare_t_d}. As shown, due to the phoneme-viseme alignment ambiguity issue, HuBERT \cite{hubert} does not align /d/ and /t/ to a similar lip shape. Conversely, due to the design of cross-modal alignment, our approach effectively addresses this issue.

\noindent\textbf{User Study.} We conducted a user study to assess the quality of the synthesized videos effectively. We sampled 20 video clips from the quantitative evaluation and invited 12 volunteers to participate in the study. We used the mean opinion score (MOS) as the metric. Volunteers were asked to rate the synthesized videos on three aspects: 1) Lip-sync Accuracy, 2) Image Quality, and 3) Video Realness. The average scores for each method are presented in Table \ref{user study}. As shown, our method significantly outperforms the comparison methods in terms of Lip-sync Accuracy, indicating the effectiveness of the cross-modal alignment framework.

\subsection{Model Inference Efficiency}
We present the inference efficiency of different models in Table \ref{eff}. The experimental results show that the proposed PASE achieves superior hardware efficiency while maintaining competitive inference speed. Specifically, PASE attains a 8.8× parameter reduction compared to Wav2vec 2.0 (37.7M vs 317M) with only 38.4\% longer inference time (4.72s vs 3.41s), while significantly reducing GPU memory consumption by 57.3\% (2051.5MB vs 4805.3MB). Compared to the lightweight Whisper, PASE shows comparable computational efficiency (4.72s vs 5.03s). These metrics collectively verify our design's effectiveness in balancing computational efficiency and phoneme-viseme alignment accuracy.

\begin{table}[t]
\caption{The results of the inference efficiency.}
  \centering
    \scalebox{0.85}{\begin{tabular}{cccc}
    \toprule
    \textbf{Model} & \textbf{Inference Time (s)} & \textbf{Params(M)} & \textbf{GPU Usage (MB)} \\
    \midrule
    \textbf{HuBERT \cite{hubert}} & 3.36 $\pm$ 0.3 & 316   & 4348.2 \\
    \textbf{DeepSpeech \cite{deepspeech}} & 191.44 $\pm$ 3.7 & -     & 523.2 \\
    \textbf{Wav2vec 2.0 \cite{wav2vec}} & 3.41 $\pm$ 0.3  & 317   & 4805.3 \\
    \textbf{Whisper \cite{whisper}} & 5.03 $\pm$ 0.2 & 19.8  & 1032.7 \\
    \textbf{PASE(Ours)} & 4.72 $\pm$ 0.2 & 37.7  & 2051.5 \\
    \bottomrule
    \end{tabular}%
    }
    \label{eff}
\end{table}%

\subsection{Ablation Study}\label{ablation study}
\begin{table}[t]
    \caption{The results of ablation study on speech spectrogram and modeling. We highlight the \textbf{best} and \underline{second best} results.}
    \centering
\scalebox{0.75}{\begin{tabular}{c|ccc|ccc}
    \toprule
    & & \textbf{NeRF}& & &\textbf{3DGS} & \\
    \midrule
          & \textbf{PSNR↑} & \textbf{LPIPS↓} & \textbf{LMD↓} & \textbf{PSNR↑} & \textbf{LPIPS↓} & \textbf{LMD↓} \\
    \midrule
    \textbf{STFT+GRU(PASE)} & \textbf{32.2302} & \textbf{0.0399} & \textbf{2.8725} & \textbf{31.0176} & \underline{0.0401} & \textbf{2.6836} \\
    \midrule
    \textbf{Mel+GRU} & 31.7962 & 0.0419 & 3.1194 & 30.4452 & 0.0425 & 2.8709 \\
    \textbf{Mel+CNN} & 31.7617 & 0.0417 & 3.0929 & 30.6923 & 0.0402 & 2.9109 \\
    \textbf{STFT+CNN} & \underline{31.8757} & \underline{0.0413} & \underline{3.0089} & \underline{30.7702} & \textbf{0.0400} & \underline{2.7432} \\
    \bottomrule
    \end{tabular}%
    }
    \label{ablation}
\end{table}
To evaluate the effectiveness of the PASE framework, we conducted ablation experiments from two dimensions: speech spectrograms and modeling. The experimental results are presented in Table \ref{ablation}. As shown, the combination of STFT spectrogram and GRU significantly outperforms other variants in both video fidelity and lip sync accuracy, highlighting the necessity of the collaborative design of modules in PASE.

\noindent\textbf{Speech Spectrogram Comparison.} The STFT spectrogram exhibits a significant advantage over the Mel spectrogram. Under the NeRF rendering model, the LMD of STFT+GRU is reduced by 8.0\% compared to Mel+GRU. This is because STFT’s linear frequency domain partitioning (0-8kHz full frequency range) preserves more frequency details. For example, STFT can retain high-frequency features of fricatives like /s/ and /\textipa{S}/ (4-8kHz), while the Mel (dominated by 0-4kHz) leads to the loss of such critical visual information. STFT spectrogram achieves the best or second-best performance across all variants, indicating that feature details are crucial for improving the performance of the rendering model.

\noindent\textbf{Modeling Comparison.} The temporal characteristics of GRU offer an improvement over the static CNN encoder. Under the 3DGS rendering model, the LMD of STFT+GRU is reduced by 2.2\% compared to STFT+CNN. The update gate mechanism of GRU can adaptively adjust the granularity of temporal modeling, while CNN’s fixed receptive field fails to capture dynamic changes at the phoneme level. The experiments show that joint optimization of temporal modeling and full-band spectrograms is crucial for improving lip sync accuracy.


\section{Conclusion}

In this paper, we addressed the challenge of phoneme–viseme alignment ambiguity in talking head synthesis. To this end, we proposed PASE (Phoneme-Aware Speech Encoder), a novel and generalizable speech representation model that explicitly bridges the gap between phonemes and visemes. By introducing phoneme embeddings as alignment anchors and incorporating a contrastive alignment module, PASE effectively enhances the discriminability and consistency of audio–visual correspondence. Moreover, the prediction and reconstruction task further improves robustness under noisy environments and partial modality absence. Comprehensive experiments on both NeRF- and 3DGS-based rendering frameworks demonstrate that PASE significantly improves lip sync accuracy, achieving 13.7\% and 14.2\% gains over conventional acoustic features. Furthermore, PASE serves as a plug-and-play speech encoder, seamlessly integrated into existing talking head pipelines to improve the lip sync accuracy without architectural modifications.

\bibliographystyle{IEEEtran}
\bibliography{main.bib}

\end{document}